# Estimating Random Delays in Modbus Network Using Experiments and General Linear Regression Neural Networks with Genetic Algorithm Smoothing

B. Sreram[1], F. Bounapane[2], B. Subathra[3], Seshadhri Srinivasan[4]

[1] Department of Information Technology, Kalasalingam University, Krishnankoil (via), Srivilliputtur, Tamilnadu
`bsreram85@gmail.com`

[2] Department of Engineering, University of Federico II, Naples, Italy
`furio.buonopane@gmail.com`

[3] Department of Instrumentation and Control Engineering, Kalasalingam University, Krishnankoil (via), Srivilliputtur, Tamilnadu
clk0602@nitt.edu
`clk0602@gmail.com`

[4] International Research Center, Kalasalingam University, Krishnankoil (via), Srivilliputtur, Tamilnadu
`cpscourse@klu.ac.in`

**Abstract.** Time-varying delays adversely affect the performance of networked control systems (NCS) and in the worst-case can destabilize the entire system. Therefore, modelling network delays is important for designing NCS. However, modelling time-varying delays is challenging because of their dependence on multiple parameters such as length, contention, connected devices, protocol employed, and channel loading. Further, these multiple parameters are inherently random and delays vary in a non-linear fashion with respect to time. This makes estimating random delays challenging. This investigation presents a methodology to model delays in NCS using experiments and general regression neural network (GRNN) due to their ability to capture non-linear relationship. To compute the optimal smoothing parameter that computes the best estimates, genetic algorithm is used. The objective of the genetic algorithm is to compute the optimal smoothing parameter that minimizes the mean absolute percentage error (MAPE). Our results illustrate that the resulting GRNN is able to predict the delays with less than 3% error. The proposed delay model gives a framework to design compensation schemes for NCS subjected to time-varying delays.

## 1  Introduction

Networked Control Systems (NCS) use networks for information exchange among control components. Network proliferation into control loops has enabled many novel applications (see, [1]-[6] and references therein) with distinct advantages that were not realizable with traditional hard wired systems. In spite of such advantages, design and analysis of NCS has been difficult mainly due to the time-varying delays introduced by communication channels. Such delays are potential enough to adversely affect NCS performance and in the worst case can lead to instability. Therefore, it becomes imperative to model delays and capture their influence on NCS. However, this is not straight forward due to the dependency of delays on numerous network parameters such as length, contention ratio, connected devices, channel loading and network protocol that are inherently random. Further, delay variations with time are non-linear and capturing these variations with conventional models is complex. Therefore, new models that capture the influence of various factors, non-linear and time-varying behavior of delays are required for designing NCS. Objective of this investigation is to develop one such model for estimating delays that can be used in designing NCS.

Modelling random delays in communication channels for designing NCS controllers has been investigated by researchers in the past and many approaches have been proposed. The available methods can be broadly classified into two categories: *(i)* deterministic and *(ii)* stochastic delays. The first approach tries to capture the time-varying delays to be deterministic by estimating the worst-case bounds. For instance, the investigation in [7], time varying delays is modelled using buffers that reflect the delays in the channel. However, deterministic delay models in literature are conservative as they consider worst-case delays in modelling. Recently, stochastic models have gained significant attention. Time-varying delays have been modeled to be stochastic using empirical distribution [8], Markov chain [9], Markov Chain Monte Carlo [10] and Hidden Markov Model (HMM) [11]. Empirical models capture delays using stochastic distribution (For e.g. Gaussian [12], [22]) and have been used in adaptive controller design. The authors reported that the performance of controllers could be significantly improved provided an accurate estimation of delays. Except for the HMM model, the other approaches can model delays considering any one of the channel condition. Further, they can be used to design only stochastic controllers that are difficult to adapt in industries. Therefore, new models that can consider influences of multiple factors, non-linear and time-varying behavior of delays are required for designing controllers for NCS. To our best knowledge, current delay models cannot model delays considering the above mentioned factors except for the method proposed in [23], wherein data-mining techniques have been used to model the delays. Further, the available delay data from time-stampings is not used in these models. To overcome these research gaps, this investigation proposes to model delays using artificial neural networks due to their ability to model complex non-linear and time-varying phenomenon (For e.g. see [13]-[14]). Further, experimental data from networks can be integrated into the model while employing the ANN based estimation techniques. In particular, this investigation uses the general regression neural network (GRNN) due to its ability to model complex, multivariate and time-varying process (see, [15]-[16] and references therein). Further, as the accuracy of estimates with GRNN depends on the smoothing parameter, this investigation uses genetic algorithm (GA) evolutionary optimization algorithm for computing the optimal smoothing parameter with the objective of reducing the mean absolute percentage error (MAPE). The resulting delay model gives more accurate estimates than the conventional GRNN.

Main contributions of this investigation are: *(i)* Experiments to model time-varying delays, *(ii)* GRNN model for time-varying delays and *(iii)* improvement in GRNN performance using genetic algorithm.

The paper is organized into five sections. Section II, presents the experimental study with MODBUS over TCP/IP network. The GRNN model and GA optimization algorithm for computing the optimal smoothing parameter is described in Section III. Comparison of the actual delays and the estimated ones using GRNN with and without RCGA smoothing are presented in section IV. Conclusions are drawn from the obtained results in section V.

## 2 Experiments to Model Time-Varying Delays

The first step to model time-varying delays in NCS is the collection of data containing the network conditions and delays. This data is the input to the GRNN for estimating the delays. The data will be used in two modes, training and testing. During the training mode, the GRNN is adjusted based on the inputs until sufficient accuracy is obtained. Once trained the GRNN produces delay estimates, when presented with the network conditions. Therefore, collection of data is pivotal to the accuracy of delay estimates. Further, the experiments should capture the various conditions envisaged during the NCS operation. This requires changing network conditions and recording delays. Then we use these conditions to generate the delay estimates.

This investigation tries to model the delays in an industrial network. Therefore, we select MODBUS over TCP/IP as the candidate network whose delays will be modeled using GRNN. The selection is motivated by the wide application of the MODBUS in industries [18]. The experimental prototype is shown in Figure 1. It consists of a Master Controller and number of slave controllers, the Modbus over TCP/IP network is used to connect the slave controllers. The network loading, length, channel contention by varying the devices and the number of rungs of the programmable logic controller (PLC) logic was changed. The delays for variations in these parameters were recorded using simple program in CoDeSys [19] following IEC 61131. The delays and the network conditions are the input to the GRNN model, while the delay estimates are the output. The GRNN uses the input samples, and estimates the time-varying delays. This experimentation procedure leads to determination of delay samples that is essential for estimating the delays.

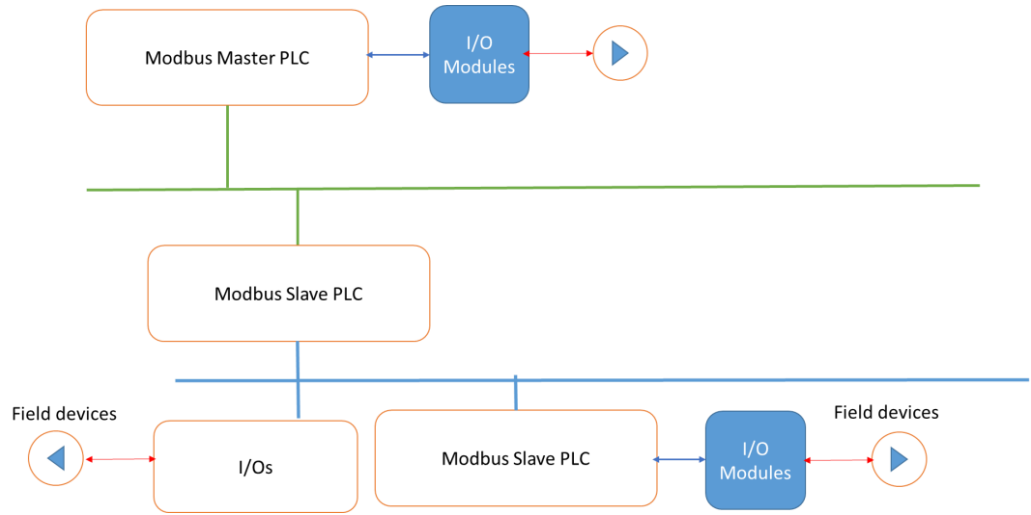

**Fig. 1.** Experiment on Modbus over TCP/IP

## 3. Genetic Algorithm Tuned General Regression Neural Network

The delay model proposed in this investigation uses the general regression neural network (GRNN); a feed-forward neural network developed by Specht [20]. The GRNN is the neural version of Nadarya-Watson kernel regression proposed in [21]. It has its basics on well developed statistical principles and converges asymptotically with an increasing number of samples to the optimal regression surface. The GRNN was selected for estimating the delays due to its ability to estimate non-linear, multi-parameter and time-varying process. A few problems with the existing approaches in using GRNN are the selection of network structure, and tuning parameters (e.g. learning rate) given by the algorithm developer leading to uncertainties and variations in the output. On the other hand, the GRNN proposed in this investigation uses the genetic algorithm to tune the spread. Therefore, ambiguities in setting up the smoothing parameter is eliminated using optimization as a decision support tool. The use of GA is motivated due to its ability to reach global optimum.

### 3.1 GRNN

GRNN is a non-linear regression tool and it is preferred over BPNN as it requires less number of samples to converge. The structure of GRNN is different from ANN and it consists of four layers: *(i)* input, *(ii)* pattern, *(iii)* summation, and *(iv)* output. The schematic of GRNN is shown in Figure 2. Unlike BPNN, GRNN employs conditional exception to calculate the regression of dependent variable y on independent variable x.

Let X is measured value of random variable x. The joint probability density function (PDF) of measurement value X and the dependent variable y is denoted by $f(X, y)$. Then the conditional mean $E[y \mid x]$ and the regression $Y(X)$ for a given X can be calculated by:

$$\hat{Y}(X) = E[y \mid x] = \frac{\int_{-\infty}^{\infty} y\, f(X, y) dy}{\int_{-\infty}^{\infty} dy\, f(X, y)} \tag{1}$$

If $f(X, y)$ is not known, it must usually be estimated from a sample of observations of x and y. An estimate of the non-conditional mean, denoted by

$$\hat{Y}(X, y) = \frac{\sum_{j=1}^{n} Y' \exp(-\frac{D^2}{2\sigma^2})}{\sum_{i=1}^{n} \exp(-\frac{D^2}{2\sigma^2})} \tag{2}$$

where σ denotes the smoothing parameter of the GRNN and, $D_i^2$ is given by,

$$D_i^2 = (X - X')^T (X - X') \tag{3}$$

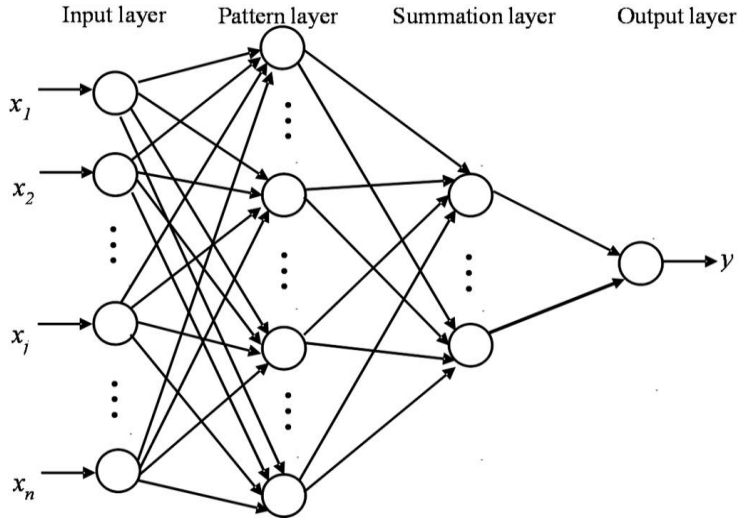

**Fig. 2.** Architecture of GRNN

Here, equation (2) can be directly used for problems with numerical data and the smoothing parameter determines the shape of the estimated density to vary between multivariate Gaussian to non-Gaussian shapes. Therefore, the GRNN performance varies considerably depending on the smoothing factor. Therefore, determining optimal value of σ is important. This study uses GA algorithm to compute the optimal parameter.

### 3.2 Genetic Algorithm

Genetic algorithms are evolutionary optimization algorithm that uses ideas of natural selection and genetics. A random search is used to solve an optimization problem. The algorithm starts with a random set of solution in the search space called population, solutions from one population is used to generate next set population that are better than the older one using genetic operators such as cross-over and mutation. Initially, a random population of chromosomes are selected. Then, the fitness function of the individual chromosomes are evaluated. A selection step is then executed. Selection involves selecting two parent chromosomes with high fitness function. The crossover operation is done on the parent chromosomes to produce new offspring. The new offspring is mutated and it is placed in new population. This new population is further used to run the algorithm. This sequence continues until the best solution is obtained.

### 3.3. Genetic Algorithm Optimized GRNN

The optimal value of σ is obtained while the output error of the GRNN is smallest. To obtain the optimal value, chromosomes to assign random values to each smoothing parameter generated within a given spread range is generated during the first step of the GA. In the second step, the fitness given by (4) is evaluated. Then the optimization of the GRNN using the GA is depicted in Figure 4.

$$J = \frac{1}{MAPE} \qquad (4)$$

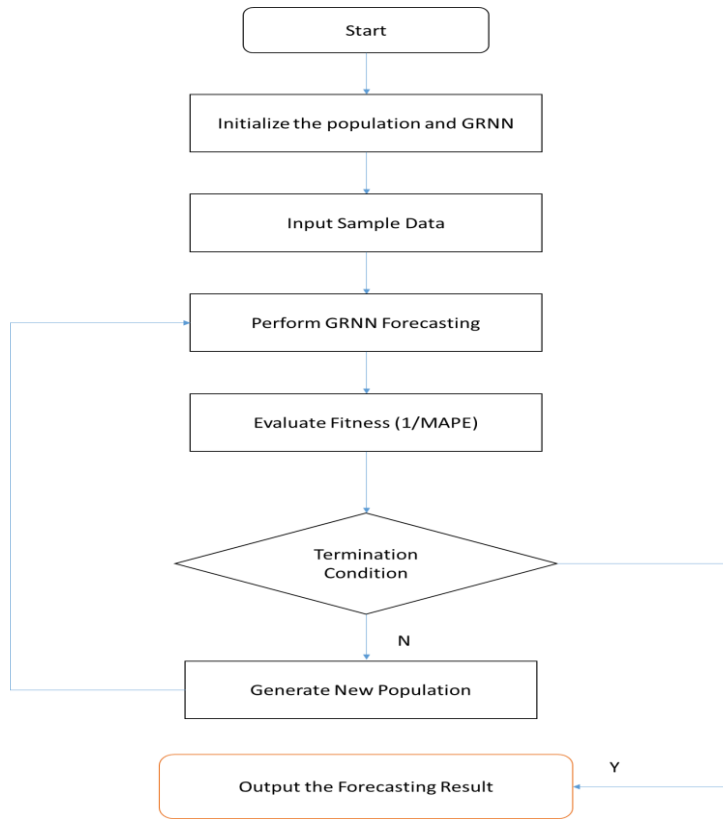

**Fig. 4.** Flowchart of GA optimized GRNN

## 4   Results and Discussions

This section presents the results of the experiment and the delay estimation. Figure 5 shows the input to the GRNN, which are the network conditions such as loading, length, contention ratio, and connected devices along with the delay samples collected from experiments. The delays are used to train the GRNN during the learning phase. Once trained the GRNN, when presented with the network condition provides the estimates of the delays.

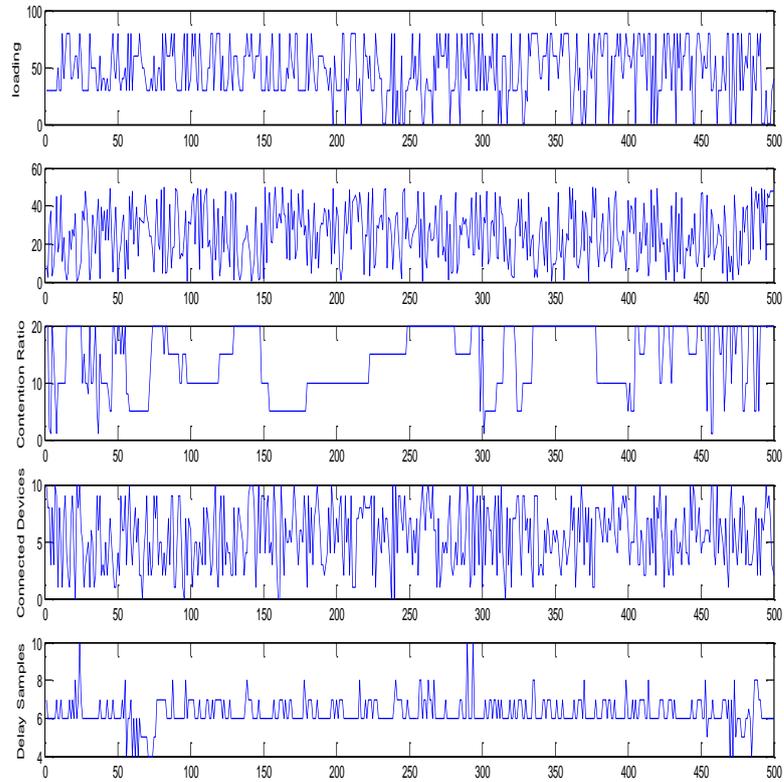

**Fig. 5.** Input to the GRNN network

The delays estimated with GRNN without optimizing the network using genetic algorithm is shown in Figure 6. The delay samples used during the validation, the delay estimates and error are shown. Our results indicate that the GRNN without GA optimization results in an estimation error of around 16% MAPE. The smoothing factor was selected ad-hoc and therefore, the estimation accuracy is low.

The estimated delay and actual delay samples with GA optimized GRNN and the error are shown in Figure 7. Our results indicate that with GA optimization of the smoothing parameter, the performance of GRNN improves significantly showing a MAPE of around 3-4% which is a significant improvement in accuracy, considering the time-varying and non-linear nature of delays.

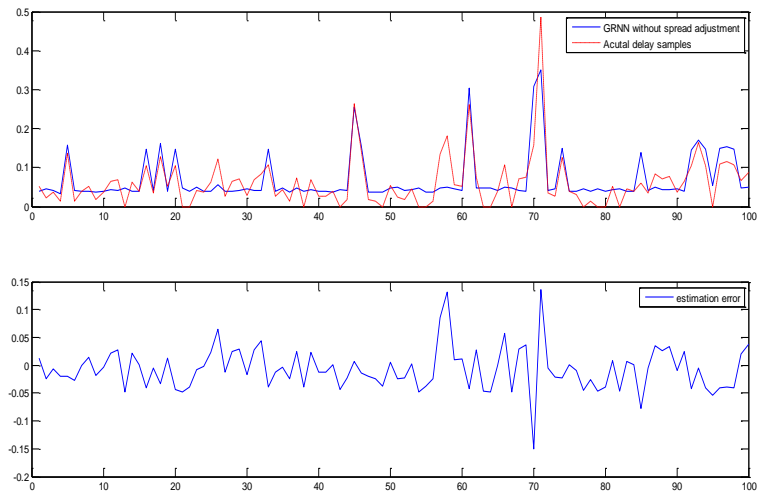

**Fig. 6.** Actual delay samples, delay estimates, and estimation error with GRNN

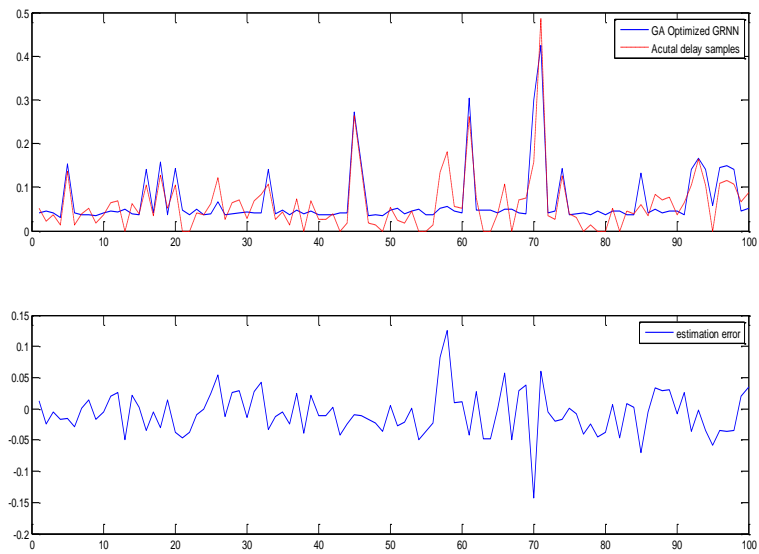

**Fig. 7.** Actual delay samples, delay estimates, and estimation error with GA optimized GRNN

## 5 Conclusion

This investigation presented a new modelling approach for time-varying delays in NCS that estimated time-varying network delays considering various factors such as length. Channel loading, network protocol, contention for the channel, and channel loading. The delays were recorded for various network conditions and were used to train the GRNN. To obtain the delay samples experiments were conducted on Modbus over TCP/IP network in an industry. The output of the GRNN is the estimated delays. The delays samples obtained from experiment using time-stamps and the network conditions recorded are used to train the GRNN. As the accuracy of the GRNN depends on the smoothing parameter, GA was used to obtain its optimal value that reduces the MAPE. Once trained, GRNN produced delay estimates based on observed network conditions. Our results show that the GRNN model for delay can be used predict delay with an accuracy of 2-3% MAPE, which is a significant accuracy considering the time-varying and non-linear delays. The delay samples obtained from GRNN models can be used design controllers and design of delay compensation schemes. Further, the proposed model gives a framework to model delays considering various channel conditions, such a model is not been reported in literature. Use of the delay models in designing controllers for NCS and delay compensation schemes are future course of this investigation.